\documentclass[%
 reprint,
 amsmath,amssymb,
 aps,pra,
 floatfix,
]{revtex4-2}

\usepackage{graphicx}
\usepackage{dcolumn}
\usepackage{bm}

\usepackage[utf8]{inputenc}
\usepackage[T1]{fontenc}
\usepackage{mathptmx}
\usepackage{etoolbox}
\usepackage{color}
\usepackage{amsmath}
\usepackage{amsfonts, amssymb}

\newcommand{\br}{\mathbf{r}}

\newcommand{\by}{\mathbf{y}}

\begin{document}

\title{Image charge effects under metal and dielectric boundary conditions}

\author{Tingtao Zhou}
\email{edmondztt@gmail.com}
\affiliation{ 
Divisions of Engineering and Applied Sciences and Chemistry and Chemical Engineering, California Institute of Technology,
Pasadena, California 91125, USA}

\author{Dorian Bruch}
\affiliation{
Division of Chemistry and Chemical Engineering, California Institute of Technology,
Pasadena, California 91125, USA}

\author{Zhen-Gang Wang}
\email{zgw@cheme.caltech.edu}
\affiliation{
Division of Chemistry and Chemical Engineering, California Institute of Technology,
Pasadena, California 91125, USA}%

\date{\today}

\begin{abstract}
Image charge effect is a fundamental problem in electrostatics. However, a proper treatment at the continuum level for many-ion systems, such as electrolyte solutions or ionic liquids, remains an open theoretical question. Here, we demonstrate and systematically compare the image charge effects under metal and dielectric boundary conditions (BCs), based on a renormalized Gaussian-fluctuation theory. Our calculations for a simple 1:1 symmetric electrolyte in the point-charge approximation show that the double-layer structure, capacitance, and interaction forces between like-charged plates depend strongly on the types of boundaries, even in the weak-coupling regime. Like-charge attraction is predicted for both metal and dielectric BCs. Finally, we comment on the effects of a dielectrically-saturated solvent layer on the metal surface. We provide these results to serve as a baseline for comparison with more realistic molecular dynamics simulations and experiments.
\end{abstract}

\maketitle

\section{\label{sec:intro}Introduction}

Image charge (IC) interaction has wide implications in electrostatics problems with interfaces.
A classic example is the study on surface tension of electrolyte solutions pioneered by Wagner~\cite{wagner1924oberflachenspannung} and Onsager and Samaras~\cite{onsager1934surface}, where the IC repulsion at air/water interface was considered responsible for the increase in the surface tension due to added salt ions. Monte Carlo simulations~\cite{naji2005electrostatic,bakhshandeh2011weak} and an integral equation theory based on hypernetted chain approximation~\cite{kjellander1984correlation} have shown ion depletion near a weakly charged dielectric interface, while ion accumulation near a metal interface is expected. Nanoscale supercapacitors have received recent interest for nanoscale device applications~\cite{lemay2013single,babel2018impedance,han2019nanoscale} and may be used to probe material characteristics at the molecular level~\cite{li2010nanogap}. These nanoscale devices  require proper treatment of IC effects. IC interaction also affects the study of capacitive desalination~\cite{biesheuvel2014attractive}, capillary freezing of ionic liquids~\cite{comtet2017nanoscale}, surface adsorption of ions~\cite{chao2020effects}, ion transport in charged nano-porous materials~\cite{zhou2020freezing}, and may even play a role in planet formation~\cite{steinpilz2020electrical}. 

The Poisson--Boltzmann (PB) theory has been widely accepted as the leading order theory for the weak-coupling regime of electrostatics~\cite{naji2005electrostatic,netz2000beyond,kanduvc2007electrostatic,podgornik1989electrostatic,naji2013perspective}, but it does not include ion-correlation or IC effect. Theoretically, a primitive treatment is to explicitly construct the image charge of a single ion and modify the free energy at the interface~\cite{onsager1934surface}. In field-theoretic formulations, perturbative loop-expansion treatments have been used to derive corrections to the PB theory~\cite{kanduvc2007electrostatic,lau2020enhancement,solis2021surface} to include IC interactions. However, by analysis of a single point charge approaching a bounding interface, one can show that the image charge interaction diverges as the charge approaches the interface ~\cite{wang2014continuous}. For a dielectric interface, this singular behavior results in a singular boundary layer of the electrolyte close to the interface, eluding regular perturbation methods~\cite{netz2000beyond}. Ref.~\onlinecite{wang2010fluctuation} derived a renormalized field-theory framework with implicit solvent. There, the solvent is modeled as a continuum with an effective dielectric constant. Sharp dielectric interfaces have been studied using this framework~\cite{wang2013effects,wang2015theoretical} and lattice Monte Carlo simulations~\cite{jiang2018improved} have shown good agreement with the predictions from Ref.~\onlinecite{wang2013effects}. Other field-theoretic treatments including IC interactions usually introduce an arbitrary cutoff to avoid divergence of self-energy~\cite{podgornik1988inhomogeneous,dean2004field}.
For the metal boundary, several techniques for Molecular Dynamics (MD) simulations have been devised to account for the IC interactions, such as iterative methods~\cite{siepmann1998influence,reed2007electrochemical,limmer2013charge,nguyen2019incorporating} or explicit construction of image charges~\cite{lin2009image,gan2011multiple,xu2013electrostatic,dos2016simulations,dwelle2019constant}, and ion accumulation on the metal surfaces has been observed~\cite{dos2017simulations}. Simulations using the method of periodic Green's functions have also been constructed for both dielectric and metal boundaries~\cite{dos2017simulations}. In all of these simulations, the solvent is treated implicitly where the polarization of the solvent is modeled by a bulk dielectric constant. However, as found in Ref.~\onlinecite{son2021image}, IC interaction is strongly affected by the polarizability of both solvent molecules and ions --- a polarizable solvent even cancels out some effects of the IC, which is neglected in models with implicit solvent background.

IC effects are known to be important for electric double layer capacitors. For example, experimental evidence shows that surface polarizability leads to a much higher differential capacitance than expected~\cite{mariappan2010electrode}, and IC interaction may even lead to a phase transition at the electrolyte-metal interface~\cite{rotenberg2015structural,lee2016ion}.
For room-temperature ionic liquids, MD simulations and 1D model analysis~\cite{kondrat2011superionic,kondrat2014single} showed importance of IC effects on capacitance curves. For lower bulk salt concentrations, recent MD simulations~\cite{son2021image} have shown significant enhancement of charge separation in a nano-supercapacitor with conducting electrodes. MD simulations also showed~\cite{bagchi2020surface} non-trivial properties of double-layer structures of polyelectrolytes due to surface polarization for both dielectric and metal boundaries.

IC effects also influence colloidal interactions~\cite{israelachvili2015intermolecular,belloni2000colloidal}, which play a critical role in many soft matter systems. The classic DLVO theory~\cite{verwey1948theory} predicts only screened repulsive electrostatic interactions. More recent efforts extending the PB theory, such as the renormalized jellium model~\cite{trizac2004renormalized,pianegonda2007renormalized,colla2010renormalized,colla2012equation,de2016renormalized}, result in a renormalized charge but qualitatively same repulsive tendency. In fact, Trizac has shown that generally local density approximations lead to repulsive pair potentials~\cite{trizac2000effective}. 
Nevertheless, like-charged colloidal attraction have been reported~\cite{kjellander1990theoretical,bowen1998long,gelbart2000dna} and has wide implications. For example, during cement setting, attraction between like-charged surfaces may allow the cohesion strength to increase beyond the order of magnitude of capillary stress~\cite{zhou2019capillary,zhou2019multiscale,monfared2020effect} or van der Waals interaction. Without systematic considerations of IC effects, the like-charge attraction has been mostly attributed to ion-ion correlations and fluctuations~\cite{kjellander1984correlation,kjellander1988double,pellenq1997electrostatic,pincus1998charge,levin1999like,linse1999electrostatic,netz2000beyond,netz2001electrostatistics,pellenq2004does}. These correlation effects are generally thought to become significant at the strong-coupling limit~\cite{moreira2000strong,moreira2001binding,vsamaj2011counterions,vsamaj2011wigner}, typically from high surface charge density or multi-valent ions. 
In the weak-coupling regime, Ref.~\cite{wang2013effects} has shown like-charge attraction due to IC-induced ion depletion between dielectric plates. For 1:1 electrolyte with metal boundary conditions and fixed surface charge density, dos Santos \& Levin~\cite{dos2019like} concluded that attraction between like-charged particles can happen for spheres even at the level of the PB theory.

In reality, IC interaction can be further complicated by various other factors. These interfacial effects are often ion-specific~\cite{jungwirth2006specific}, such as the famous Hofmeister series~\cite{hofmeister1888lehre} for protein salt-in/salt-out. The hydration shell structure of the ions also changes when they are adsorbed onto a metal surface, which affects their hydration energy. In addition, more complex and unexpected behaviors of the differential capacitance have been reported in MD simulations with polyelectrolytes~\cite{bagchi2020surface}. In the case of a free interface, the IC interaction is also coupled with surface capillary waves~\cite{otten2012elucidating,stern2013thermodynamics,vaikuntanathan2014putting}.
These complications are beyond the scope of this work.

In this paper, we provide a simple theoretical picture of the IC effects as a reference for further studies with more realistic molecular models for electrolytes. To this end, we treat all ions as point charges and assume an implicit solvent background. We present a systematic comparison between the metal BC, dielectric BC, and PB theory in terms of the differential capacitance of nano-capacitors and forces between like-charged plates by numerically solving the variational field theory for an incompressible electrolyte. Our calculations show that the differential capacitance depends strongly on the boundary condition (metal or dielectric) as well as the bulk salt concentration. For example, even at low bulk concentration but if the slit width is small, the metal BC leads to a ``bird-shaped" capacitance curve~\cite{cruz2018bird}. For a simple 1:1 electrolyte, we find the interplate forces change from repulsion to attraction as the separation distance decreases between dielectric plates, and stronger and long-ranged attractions between like-charged metal plates, due to the IC interaction. Moreover, to account for the distance of closest approach of real ions to a surface, we consider a thin layer of polarizable solvent on the metal surface, preventing direct contact of ions with the boundary. We show that even a very thin surface dielectric layer cancels out some IC effects, consistent with the findings in Ref.~\onlinecite{son2021image}.

\section{Theory}

\subsection{General Gaussian variational approach
\label{sec:variational-framework}}

We  start by recapitulating the field-theoretic framework as presented in Ref.~\onlinecite{wang2010fluctuation}.
Consider a system with cations (of valence $z_+$) and anions (of valence $-z_-$) in an electrolyte solution, and external charges on the boundary surfaces $e\rho_{ex}$. The solution is connected to a reservoir with bulk cation and anion concentrations $c_{+,0}$ and $c_{-,0}$, respectively. The total microscopic charge density is
\begin{equation}
e\rho(\br) = e\left(\rho_{ex}(\br) + z_+ \sum_{i+} h_+(\br-\br_+^i) - z_- \sum_{j-} h_-(\br-\br_-^j)\right)
\end{equation}
where the salt ions have a charge spread kernel, which is taken to the point charge limit $h_{\pm}(\br-\br_{\pm}^i)=\delta(\br-\br_{\pm}^i)$ throughout this work.
The Coulomb energy is
\begin{equation}
H = \frac{e^2}{2}\int d\br d\br'  \rho(\br) G_0(\br,\br') \rho(\br')
\end{equation}
where the Coulomb operator $G_0$ is defined by
\begin{equation}
- \nabla\cdot \varepsilon(\br) \nabla G_0(\br,\br') = \delta(\br-\br')
\end{equation}
The grand canonical partition function is
\begin{equation}
\begin{split}
\Xi = & \sum_{n_{+}}\sum_{n_-} \frac{e^{n_+\mu_+} e^{n_-\mu_-}}{n_+!n_-! v_+^{n_+}v_-^{n_-}}\int d\br_{i+} d\br_{j-} e^{-\beta H}  \\
= & \frac{1}{\sqrt{\det(G_0)}} \int D[\phi] e^{-\int d\br\mathcal{L}[\phi(
\br)]} 
\end{split}
\end{equation}
where the action density is
\begin{equation}
\mathcal{L}[\phi]= \frac{1}{8 \pi \ell_B} \left(\nabla \phi(\br)\right)^2 + i \rho_{ex}(\br) \phi(\br)
- \lambda_{+} e^{ -i z_{+}\phi(\br)} + \lambda_{-} e^{ i z_{-}\phi(\br)}
\label{eqn:lagrangian-density}
\end{equation}
For convenience Equation.~\eqref{eqn:lagrangian-density} and subsequent ones throughout the paper are written in dimensionless variables. The Bjerrum length is $\ell_B=e^2/(4\pi\epsilon k_B T)$. The complex-valued auxiliary field $\phi(\br)$ conjugate to the charge density $\rho(\br)$ is introduced through the standard Hubbard--Stratonovich transformation~\cite{hubbard1959calculation,stratonovich1957method} and then non-dimensionalized by $k_BT/e$. $\lambda_{\pm}=\frac{e^{\mu_{\pm}}}{v_{\pm}}$ are the bulk activities of salt ions, where $\mu_{\pm}$ are the chemical potentials and $v_{\pm}$ are the volume scales which have no thermodynamic consequences other than shifting the reference of the chemical potential; therefore, for convenience, we choose $v_{\pm}$ to be simply the volume of the ions. A renormalization of the field theory then can be derived through a variational approach by extremizing~\cite{wang2010fluctuation,frydel2015introduction} the grand free energy
\begin{equation}
W \le W_{ref} + \left< L[\phi] - L_{ref}[\phi] \right>_{ref}
\end{equation}
The auxiliary field is decomposed into 
\begin{equation}
\phi(\br) = -i\psi(\br) + \chi(\br)
\end{equation}
where the mean-field part $-i\psi(\br)$ and fluctuation parts $\chi(\br)$ are purely imaginary and real fields, respectively.
The reference action is chosen as a Gaussian form for the fluctuation part for the auxiliary field 
\begin{equation}
L_{ref} = \frac{1}{2}\int d\br d\br' \chi(\br) G^{-1}(\br,\br') \chi(\br')
\label{eqn:ref-lagrangian}
\end{equation}
The Green's function $G(\br,\br')$ is unknown and will be determined together with the mean-field $\psi(\br)$ by the extremization of $W$, which results in
\begin{align}
-\nabla\cdot \left[ \frac{1}{4 \pi \ell_B(\textbf{r})}\nabla \psi(\textbf{r}) \right] =  \rho_{ex}(\textbf{r}) + z_+ c_+ (\textbf{r}) - z_- c_- (\textbf{r}) \label{eqn:PB} \\
-\nabla\cdot \left[ \frac{1}{4 \pi \ell_B(\textbf{r})} \nabla G(\textbf{r},\textbf{r}') \right] + 2I(\br)G(\textbf{r},\textbf{r}')
= \delta(\textbf{r}-\textbf{r}') \label{eqn:green}
\end{align}

where the ionic strength is $I(\br) = \frac{1}{2}\left[z_+^2 c_+(\br)+ z_-^2 c_-(\br)\right]$, and ion concentrations are
\begin{equation}
c_{\pm}(\textbf{r}) = \lambda_{\pm}\exp\left\{
\mp z_{\pm}\psi(\textbf{r})
- u_{\pm}(\br)
\right\} 
\end{equation}
with the self energy
\begin{equation}
\label{eqn:self-energy-Green}
u_{\pm}(\br) = \frac{1}{2} z_{\pm}^2 G(\textbf{r},\textbf{r}'\rightarrow\br)
\end{equation}
Here the self-energy appears in the exponential of the Boltzmann factor of the charge distributions and it is the key feature of this framework. It includes a singular part, which is present even in a homogeneous solution without boundaries; this singular part can be regularized by the introduction of a smearing function~\cite{wang2010fluctuation}, with a spread chosen to reproduce the Born solvation energy in a bulk electrolyte solution and can be absorbed into the bulk ion chemical potentials. The remaining part of the self-energy comes from spatial variations of charge distribution, correlation energy, and the existence of boundaries and image charges. This part of the self-energy is finite inside a continuum electrolyte, except at dielectric discontinuities and boundaries.

\subsection{Incompressible electrolyte}
Even within the PB framework, ion densities near electrolyte--metal interfaces can be overestimated and exceeds the physically maximum value of close packing~\cite{cuvillier1997adsorption}, due to the unconstrained exponential dependence of ion concentration on the electric potential. 
The point-charge approximation in the variational field theory presented in the last section still exhibits divergent self energy near an interface~\cite{wang2014continuous}, resulting in complete ion depletion (divergent accumulation) near a dielectric (metal) wall.
Here we incorporate the incompressibility constraint~\cite{borukhov1997steric,borukhov2000adsorption} into the field-theoretic framework to regulate ion concentration near the boundary surfaces.
This way, the IC interaction is still calculated for point charges that can approach the interface indefinitely close, and hence the IC force may diverge as expected, but the ion concentration is regulated by a saturation value. The incompressibility condition is represented by a product of delta functions that enforces the solution density $\frac{1}{v_0}$ at every point in space, $\prod_\textbf{r}\delta\left(\hat{c}_+(\br) + \hat{c}_-(\br) + \hat{c}_s(\br) - \frac{1}{v_0} \right)$, where the number densities of cation, anion and solvent molecules are
\begin{equation}
\begin{split}
\hat{c}_+(\br) = & \sum_{i+} \delta(\br-\br_+^i) \\
\hat{c}_-(\br) = & \sum_{i-} \delta(\br-\br_-^i) \\
\hat{c}_s(\br) = & \sum_{s-} \delta(\br-\br_s^i) 
\end{split}
\end{equation}
For simplicity we have assumed that the volumes of all species (cation, anion, and solvent) are equal $v_+=v_-=v_s=v_0=(0.5~\text{nm})^3$, so that the total concentration of all species is $c_0\approx13.33$~M everywhere.

To deal with the incompressibility constraint, we represent the product of delta functions as its Fourier transform by introducing an auxiliary field $\eta(\br)$

\begin{equation}
\begin{split}
\prod_\textbf{r}\delta\left(\hat{c}_+(\br) + \hat{c}_-(\br) + \hat{c}_s(\br) - \frac{1}{v_0} \right) &= \\
\int D[\eta(\br)] & e^{i\int d\br \eta (\hat{c}_+ + \hat{c}_- + \hat{c}_s - \frac{1}{v_0})}
\end{split}
\end{equation}

where the notation $\int D[\eta(\br)]$ is the functional integral with respect to the field $\eta(\br)$. Now $\hat{c}_{\pm,s}(\br)$ still depend on the ion/solvent positions $\{\br^i\}$, which are integrated over in the grand partition function.

With this constraint incorporated into the grand partition function, extremizing the grand free energy with respect to the new auxiliary fields gives modified equations for concentration fields
\begin{align}
c_{\pm}(\textbf{r}) = & \lambda_{\pm} e^{\mp (z_{\pm}\psi(\textbf{r})) + \eta(\textbf{r}) - u_{\pm}(\textbf{r})} 
\label{eqn:c-incompress}
\\
c_s(\textbf{r}) = & \frac{e^{\mu_s}}{v_0} e^{ 
\eta(\textbf{r}) }
\label{eqn:c-solvent}
\\
\frac{1}{v_0} = & c_+(\textbf{r}) + c_-(\textbf{r}) + c_s(\textbf{r})
\label{eqn:vol-conserve}
\end{align}
where these Equations \eqref{eqn:c-incompress} and \eqref{eqn:c-solvent} are obtained from the integrand of chemical potential derivatives of the grand free energy. Note that the differential equations~\eqref{eqn:PB} -- \eqref{eqn:green} for the electric potential $\psi$ and Green's function $G$ remain the same. Hence, the treatment for incompressibility is at the mean-field level. For consistency in the comparisons throughout this paper, we impose the incompressibility condition for all cases---PB theory, metal and dielectric BCs, and for a thin dielectric surface layer on a metal wall.

\subsection{Force and capacitance between two plates}

Consider an electrolyte solution confined between two infinite parallel plates. In this work, we focus on comparing  four  cases of boundary conditions (BCs): (1) both plates are perfect metal; (2) both plates are dielectric medium with a dielectric constant $\epsilon_p=2$, lower than that of the solution, assumed to be $\epsilon_r=80$; (3) PB theory without image charge corrections; and (4) both plates are metal with a thin dielectric surface layer with a dielectric constant $\epsilon_\text{L}$, lower than that of solution. For simplicity,  cases (2) and (4) will be referred to as dielectric and layer BC from now on, respectively. The electrolyte is connected to a bulk 1:1 solution of salt ion concentrations $c_{+,0}=c_{-,0}=c_0$; hence, the calculations are in the grand canonical ensemble.

Due to the translational and rotational symmetry in the directions parallel to the plates, it is convenient to use cylindrical coordinates ($\rho,\theta,z$). The Equations~\eqref{eqn:PB},~\eqref{eqn:green} and \eqref{eqn:c-incompress} -- \eqref{eqn:vol-conserve} are simplified to a set of 1D ordinary differential equations (ODEs) along the $z$-axis perpendicular to the plates
\begin{gather}
-\frac{1}{4 \pi \ell_B} \frac{\partial^2}{\partial z^2} \psi(z) =  \rho_{ex}(z) +  c_{+} (z) -  c_-(z) \label{eqn:2metal-poisson}\\
\delta(z-z') = \frac{1}{4 \pi \ell_B}\left\{\left[k^2+\kappa(z)^2\right] - \frac{\partial^2}{\partial z^2}\right\} G(k,z,z') \label{eqn:2metal-Green} \\
c_{\pm}(z) = \lambda_{\pm}e^{
\eta(z) - u(z) \mp \psi(z)
} \label{eqn:2metal-c} \\
u(z) = u_{Born} + u_{DH} + 
\frac{1}{4\pi} \int_0^{\infty} dk \left(
k G(k,z,z) - 2\pi \ell_B
\right) \label{eqn:2metal-self} \\
\kappa(z)^2 = 4 \pi \ell_B \left[ c_+(z)+ c_-(z)\right] \label{eqn:2metal-kappa}\\
c_s(z) = \lambda_s e^{\eta(z)}\\
\frac{1}{v_0}  = c_+(z) + c_-(z) + c_s(z) \label{eqn:2metal-vol-conserve}
\end{gather}

where we use the partial Fourier transform $G(k,z,z')$ of the Green function $G(\br,\br')$: by symmetry of the geometry $G(\textbf{r},\textbf{r}')=G(\vert \rho-\rho' \vert,z,z')$. 
One then only Fourier transforms the separation $s=\vert\rho-\rho'\vert$ in the parallel direction 
\begin{equation}
G(\vert\rho-\rho'\vert,z,z') = \frac{1}{2\pi} \int_0^{\infty} k dk J_0(k s) G(k,z,z')
\end{equation}
where $J_0(x)$ is the zeroth-order Bessel function. After the partial Fourier transform, Equation~\eqref{eqn:green} becomes Equation~\eqref{eqn:2metal-Green}. Inverse Fourier transforming $G(k,z,z')$ back and taking the limit of $\rho\rightarrow\rho'$, $z\rightarrow z'$ according to Equation~\eqref{eqn:self-energy-Green} gives Equation~\eqref{eqn:2metal-self}. In Equation~\eqref{eqn:2metal-self}, the singular part of the self-energy is replaced with a constant reference energy, including the Born energy $u_{Born}=z^2 \ell_B/(2a)$ and a Debye-H\"uckel correlation term $u_{DH}=-z^2 \ell_B\kappa_B/2$, which is assumed homogeneous inside the domain and in the reservoir, and hence can be absorbed into the bulk fugacity. $\kappa_B=\sqrt{8\pi c_0 \ell_B}$ is the inverse Debye length in the bulk solution. Only the excess part (second term of RHS in Equation \eqref{eqn:2metal-self}) will be relevant for our purpose here. By setting $\eta=0$ for the reservoir, the chemical potentials are determined by the bulk concentrations
\begin{align}
& \lambda_s = \frac{1}{v_0} - 2 c_0 \\
& \lambda_{\pm}e^{-u_{\infty}} = c_{\pm,0} = c_0
\end{align}

The BCs for the ODE Eq.~\eqref{eqn:2metal-poisson} can be set by fixing the surface potential (Dirichlet) or fixing the surface charge density (Neumann), depending on the specific application. We now specify the BCs for the ODE Eq.~\eqref{eqn:2metal-Green} for the Green's function at the surfaces $z=-L/2$ and $z=L/2$. With metal plates, there is no fluctuation inside the metal, so $G(z=-L/2,z')=G(z=L/2,z')=0$ (noting that $G(\br,\br')=\left<\chi(\br)\chi(\br')\right>$ is the variance of the fluctuation field $\chi(\br)$ from its defining Equation.~\eqref{eqn:ref-lagrangian}). With dielectric plates of relative permitivity $\epsilon_p$, different from the solvent relative permitivity $\epsilon_r$, the equation can be extended to the whole space and the dielectric constant is regarded as having a discontinuity at $z=\pm L/2$ so that $\epsilon(z)=\epsilon_p$ outside $\vert z \vert > L/2$ and $\epsilon(z)=\epsilon_r$ inside $\vert z \vert < L/2$. Accounting for this jump one arrives at~\cite{wang2015theoretical} a Robin BC for $G(k,z,z')$
\begin{align}
\epsilon_r \partial_z G(k,z=\pm L/2,z') & = \mp k\epsilon_p G(k,z=\pm L/2,z')
\end{align}
For the case of a thin dielectric layer on the metal surface, the metal surface is at fixed potential, and Equations \eqref{eqn:2metal-poisson} and \eqref{eqn:2metal-Green} have Robin boundary conditions, derived in Appendix C. 

The differential capacitance $C_d=d\sigma_s/dV$ can be readily obtained by sweeping the voltage $V$, which is symmetrically applied on both plates such that $\psi(z=\pm L/2)= V$. $\sigma_s$ is the surface charge density on a plate. We also examine forces between like-charged plates by continuously varying their separation distance and computing the total grand free energy $W$.
In any case, the free energy per unit area of the system is~\cite{wang2015theoretical}
\begin{equation}
\begin{split}
W
& = \frac{1}{2}\int dz \psi(z)\left\{\rho_{ex}(z) - c_+(z) + c_-(z)\right\} \\
& + \int dz (\eta(z)-1) \left\{c_+(z) + c_-(z) + c_s(z) \right\}\\
& + \int dz I(z) \int_0^1 [G(z,z;\xi)-G(z,z)] d\xi
\end{split}
\end{equation}
for fixed surface charges $\sigma_s$, and performing a Legendre transform gives the free energy $Y = W - 2\sigma_{s}V$ for fixed surface potentials at $\psi(z=\pm L/2)=V$.
The ``charging'' Green's function $G(z,z';\xi)$ is obtained by solving Eq.~\eqref{eqn:2metal-Green} but with $\kappa(z)^2$ replaced by $\xi \kappa(z)^2$.
The force per unit area is 
\begin{equation}
\Pi = -\frac{\partial \left(\mathcal{W}(L)-\mathcal{W}(\infty)\right)}{\partial L}
\end{equation}
and $\mathcal{W}$ is $W$ or $Y$ depending on whether surface charge or surface potential is specified~\cite{bruch2021}.

\section{Numerical results and discussions}

We numerically evaluate the Equations~\eqref{eqn:2metal-poisson} -- \eqref{eqn:2metal-vol-conserve} iteratively. For the fixed voltage (Dirichlet) boundary conditions, the Poisson equation~\eqref{eqn:2metal-poisson} is solved by the Chebyshev spectral method, and we combine the Chebyshev and shooting methods to solve the fixed charge (Neumann) boundary condition. For a given pair of $k$ and $z'$ values, the Green's function is solved by a finite difference method with non-uniform grid size for accuracy. For the inverse Fourier transform, the integration over $k$ in Equation~\eqref{eqn:2metal-self} is performed with Gauss--Laguerre quadrature.

\begin{figure}[htbp!]
\centering
\includegraphics[width=0.8\columnwidth]{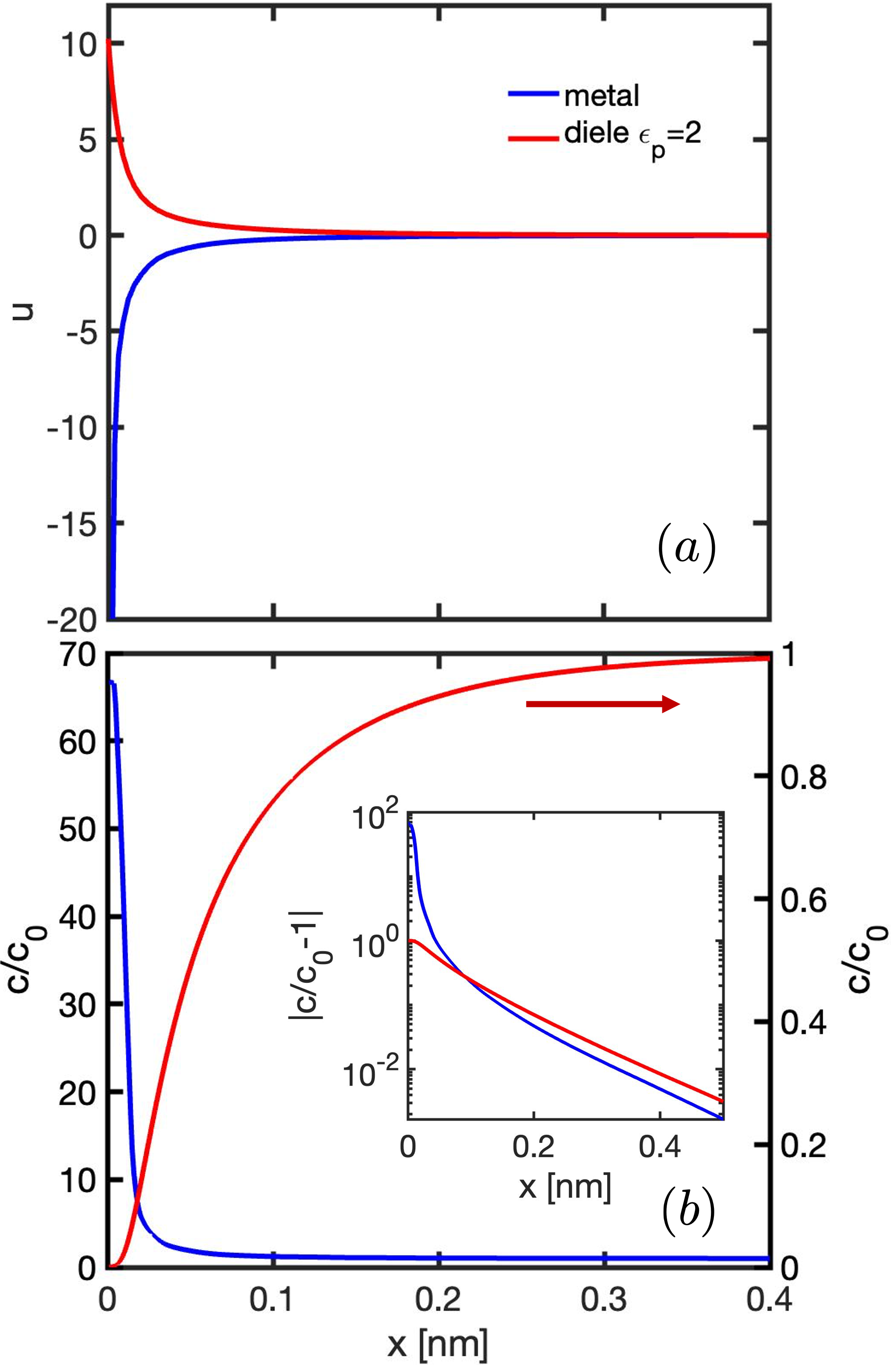}
\centering
\caption{
Charge accumulation/depletion at the planar boundary due to image charge interactions. No applied voltage or fixed surface charge on the wall. The bulk salt concentration is $c_0=0.1$~M for each case.  The dielectric constant of the implicit solvent is $\epsilon_r=80$; for the dielectric plates, $\epsilon_p=2$.  (a) Self-energy profiles corresponding to metal BC (blue) and dielectric BC (red). (b) Concentration profiles: the blue line represents the metal BC, corresponding to the left y-axis. The red line represents the dielectric BC, corresponding to the right y-axis. The inset shows the deviations from bulk concentration $c_0$ (absolute value $\vert c/c_0-1\vert$) on the same log-scale y-axis. 
\label{fig:profiles-0V-4nm}}
\end{figure}

\begin{figure}[htbp!]
\centering
\includegraphics[width=0.8\columnwidth]{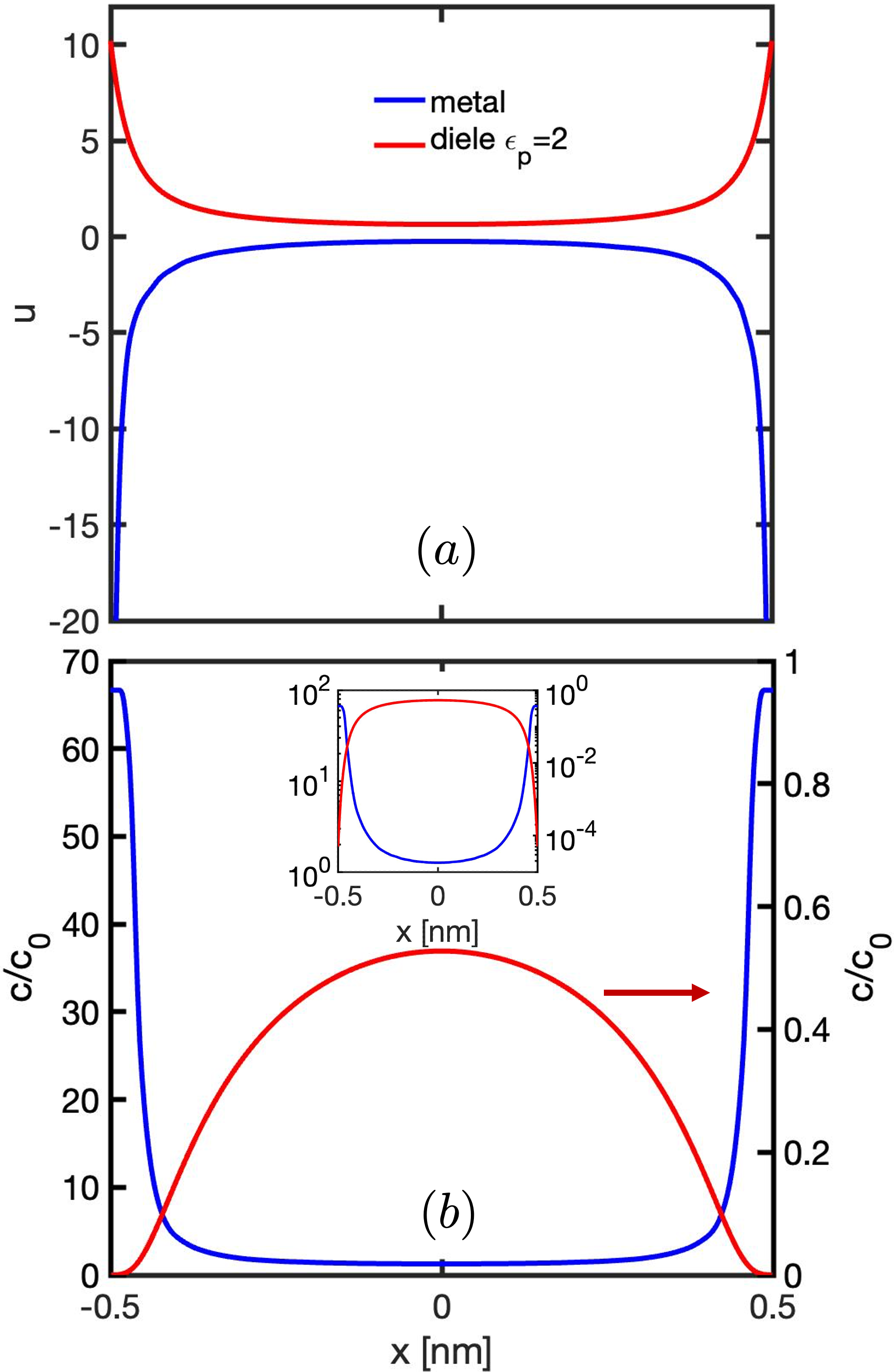}
\centering
\caption{
Finite-size effect on the charge accumulation/depletion at the boundaries due to image charge interactions without applied voltage or fixed charge. The slit pore width is $L=1$~nm and the bulk salt concentration is $c_0=0.1$~M for each case. The dielectric constant of the implicit solvent is $\epsilon_r=80$; for the dielectric plates, $\epsilon_p=2$. (a) Self-energy profiles corresponding to metal BC (blue) and dielectric BC (red). (b) Concentration profiles: the cation and anion profiles overlap for either metal or dielectric BC since there is no voltage applied. The blue line shows the metal BC and red line shows the dielectric BC. The inset shows both profiles in semilog-scale. 
\label{fig:profiles-0V-1nm}}
\end{figure}

\begin{figure}[htbp!]
\centering
\includegraphics[width=0.8\columnwidth]{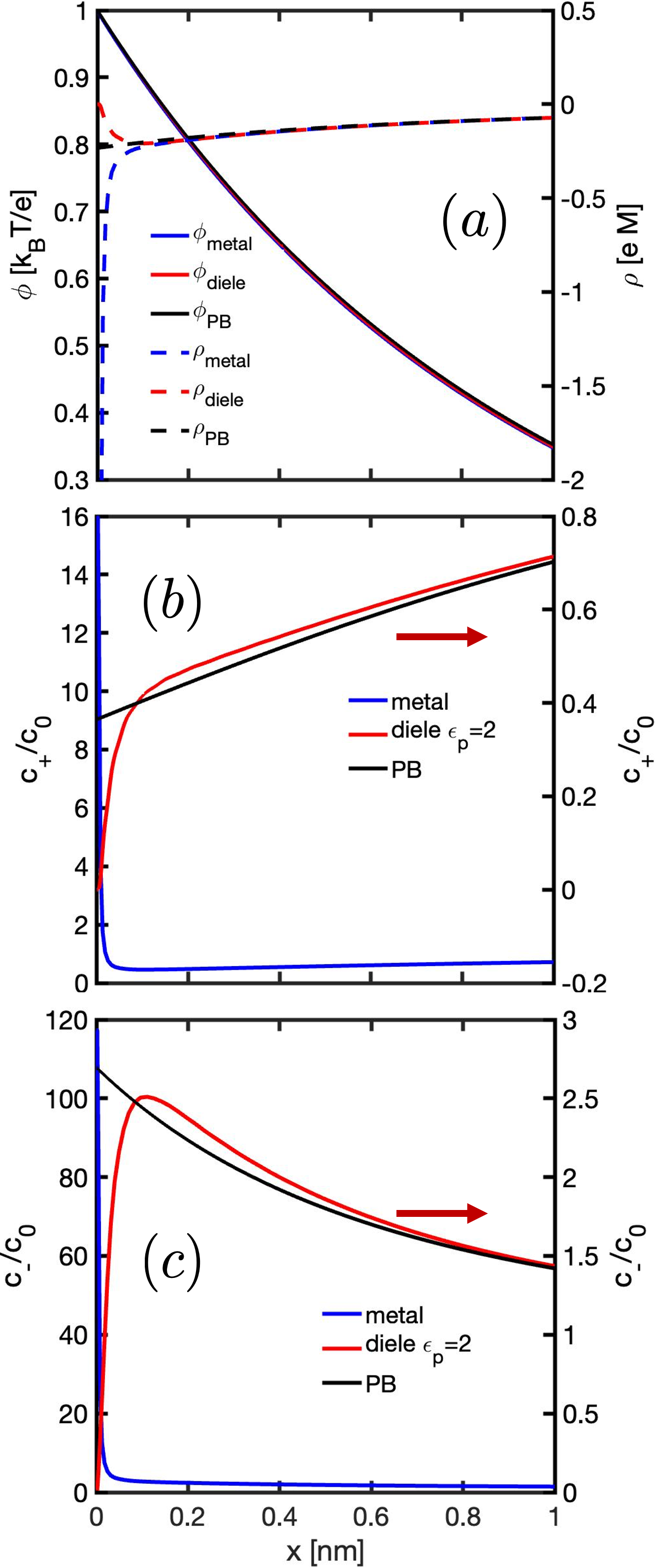}
\centering
\caption{
Single wall fixed at surface potential $V=1$. Bulk ion concentrations are $c_{0,+}=c_{0,-}=0.1$~M. To amplify the boundary layer, we zoom into the region $0<x<1$~nm. (a) Electric potential profiles for the metal (blue solid), dielectric (red solid) and PB (black dashed) BCs. (b) Cation profiles with metal BCs (blue, left-y axis), dielectric (red, right y-0axis) BCs, and PB theory (black, right y-axis). (c) Anion profiles: legends and axes are the same as in (b).
\label{fig:profiles-1V-8nm}}
\end{figure}

\begin{figure}[htbp!]
\centering
\includegraphics[width=0.82\columnwidth]{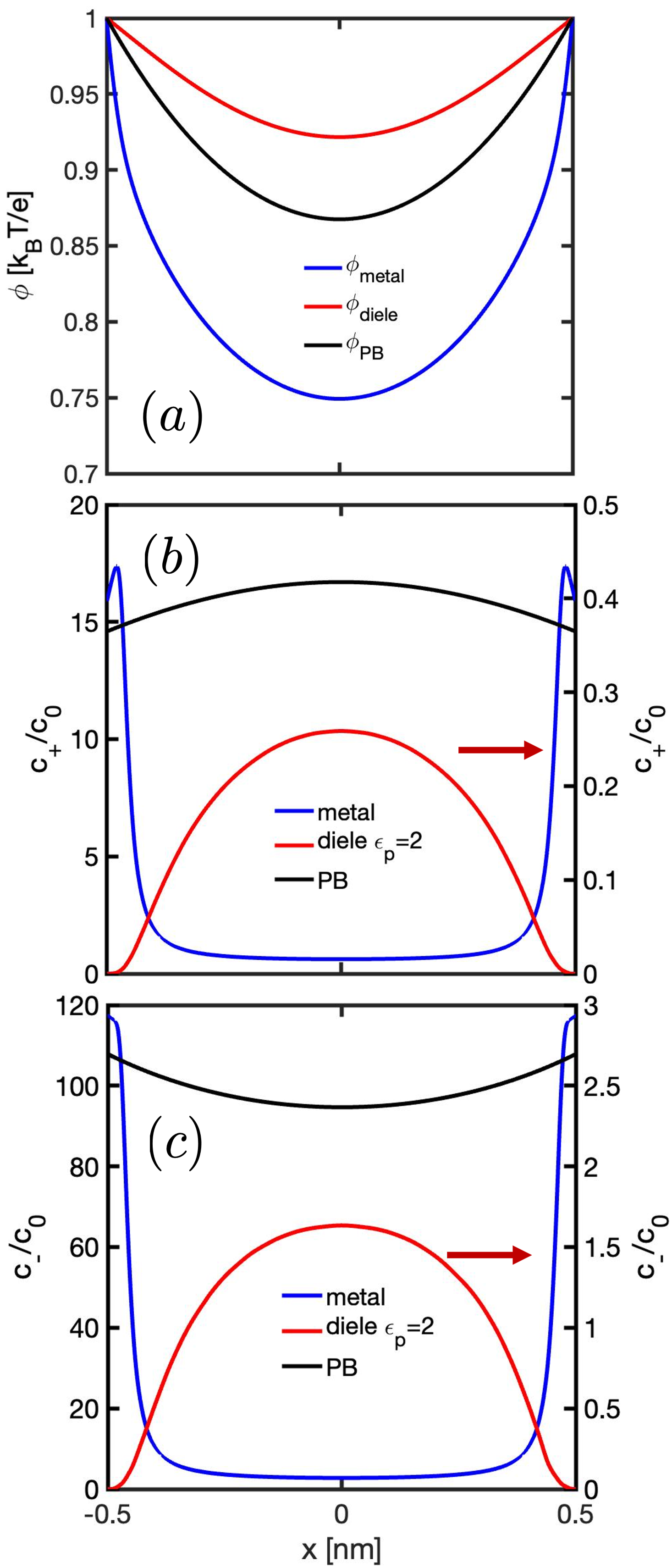}
\centering
\caption{
Symmetrically charged slit pore of 1~nm with both plates at fixed surface potentials $V_L=V_R=1$. Bulk ion concentrations are $c_{0,+}=c_{0,-}=0.1$~M.
(a) Electric potential profiles for the metal BCs (blue solid), dielectric (red solid) BCs, and PB theory (black dashed). 
(b) Cation profiles with metal BCs (blue, left y-axis), dielectric BCs (red, right y-axis) and PB theory (black, right y-axis).
(c) Anion profiles: legends and axes are the same as in (b). 
\label{fig:profiles-1V-1nm}}
\end{figure}

\begin{figure}[htbp!]
\centering
\includegraphics[width=0.78\columnwidth]{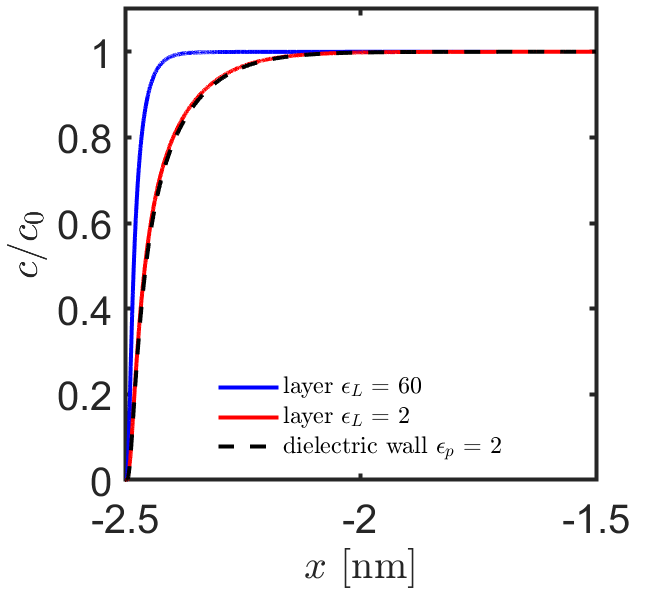}
\centering
\caption{
Ion concentration versus distance from a single metal wall at fixed surface potential $V=0$ with a dielectric surface layer. The dielectric layer has size $a = 0.1$ nm and dielectric constant $\epsilon_\text{L} = 60$ (blue) or $\epsilon_\text{L} = 2$ (red). The bulk ion concentrations are $c_{0,+} = c_{0,-} = 0.1$ M. For comparison, the ion concentration near a dielectric wall ($\epsilon_\text{p} = 2$) with no layer is shown in black. The implicit solvent has dielectric constant $\epsilon_\text{r} = 80.$\label{fig:layer_singlewall}}
\end{figure}

\begin{figure}[htbp!]
\centering
\includegraphics[width=0.85\columnwidth]{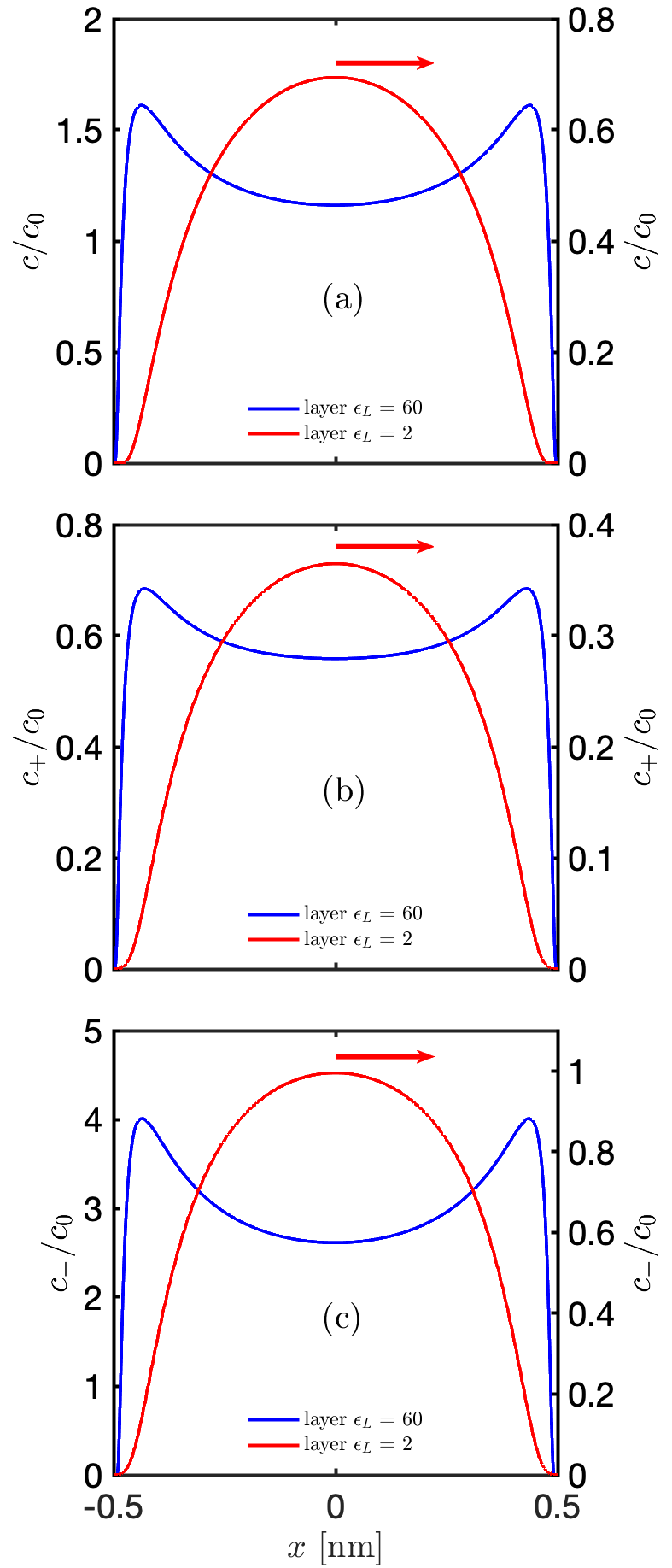}
\centering
\caption{
Ion concentration versus distance for a symmetrically charged slit pore of $L = 1$ nm. Both walls are metal with a dielectric layer of size $a = 0.1$ nm on each metal surface. The dielectric layer has dielectric constant $\epsilon_\text{L} = 60$ (blue) or $\epsilon_\text{L} = 2$ (red). The bulk ion concentrations are $c_{0,+} = c_{0,-} = 0.1$ M. The implicit solvent has dielectric constant $ \epsilon_\text{r} = 80$. (a) Applied potential $V = 0$. (b) Positive ion densities for applied potential $V = 1$. (c) Negative ion densities for applied potential $V = 1$. \label{fig:layer_den}}
\end{figure}

\subsection{Charge accumulation/depletion at the surface}

For a single point charge near a planar surface, its image charge is known to be attractive for a metal boundary and repulsive for a dielectric medium with a lower dielectric constant than that of the electrolyte. As the point charge approaches the boundary interface, its distance to the image charge decreases to 0, resulting in a divergent image charge interaction. This is reflected in the self-energy diverging at the domain boundaries as shown in Fig.~\ref{fig:profiles-0V-4nm}(a). The wall surface for both curves in Fig.~\ref{fig:profiles-0V-4nm} are fixed at $V=0$, and bulk ion concentrations are $c_{+,0}=c_{-,0}=0.1$~M. The self-energy near the surface has two contributions: (1) there are no other salt ions inside the plate, hence less correlation energy. (2) the dielectric property of the plate differs from the solution, leading to IC interactions. For dielectric BC, salt ions are completely depleted from the surface; for metal BC, the ion density saturates at the surface as shown in Fig.~\ref{fig:profiles-0V-4nm}(b). The cation and anion profiles overlap for either metal or dielectric BC since there is no applied voltage or fixed charges. Although in both cases the electric potential is constant 0 everywhere, the concentration profiles are nontrivial and nonuniform, different from the PB solution that gives trivial constant concentrations equal the bulk value of 0.1~M. Away from the boundary-layer zone, both dielectric and metal BCs show exponentially decaying deviations of ion densities from the bulk with a similar screening length, while the metal curve shows a significantly different screening length for the boundary layer zone, as shown in the inset of Fig.~\ref{fig:profiles-0V-4nm}(b). 

For two parallel plates (a slit pore) with $V=0$ on both, the double layers from the two surfaces start to overlap as the separation between the plates decreases. Hence, the middle-point self-energy and ion concentrations deviate from the bulk values as shown in Fig.~\ref{fig:profiles-0V-1nm}. For metal BC (blue line), the image charge attraction is not fully screened at the middle point so that $c(x=0)/c_0\approx1.27>1$. This is more clearly shown in the inset where the blue line does not reach $10^0$ and is not flat in the middle. For dielectric BC (red line), the image charge repulsion leads to ion depletion in the entire slit, as shown by the right y-axis of Fig.~\ref{fig:profiles-0V-1nm}(b), and the middle point concentration is significantly reduced from the bulk value.

We now examine the double layer structure when a weak but non-zero electric potential is maintained at the boundary surface(s). Fig.~\ref{fig:profiles-1V-8nm} shows the ion concentrations and electric potential profiles near a single wall fixed at $V=1$. The bulk ion concentrations are $c_{+,0}=c_{-,0}=0.1$~M. The electric potential profiles for all three boundary conditions are deceptively similar everywhere, as shown by the solid lines corresponding to the left y-axis of Fig. 3(a). However, the charge densities are different at a very short distance from the the wall due to the drastic differences in self-energy, as shown by the dashed lines corresponding to the right y-axis of Fig.~\ref{fig:profiles-1V-8nm}(a). Beyond the boundary layer, the three charge densities converge to the same outer solution. The IC interaction also leads to non-monotonic concentrations of anions for dielectric BC near the wall, as shown by the red line in Fig.~\ref{fig:profiles-1V-8nm}(c). 

For a slit pore, as its width decreases with the surface potential of both plates fixed at $V=1$, the double layers from the two surfaces start to overlap. For this case, one can see the difference between the electric potentials of the different BCs as shown in Fig.~\ref{fig:profiles-1V-1nm}(a). Overall, the concentration inside the slit pore is largest for metal BC and weakest for dielectric BC. This leads to a different magnitude of the screening effect, which mostly comes from the boundary layers. Going from the surface into the solution, for metal BC the electrostatic potential drops faster than the PB theory, whereas for dielectric BC, it drops slower than the PB theory. Compared to the single-wall case (Fig.~\ref{fig:profiles-1V-8nm}(b) and (c)), the overlapping double layers and stronger IC effects in the 1~nm pore lead to two main differences: (1) the boundary layers for both the metal and dielectric BCs become thicker; (2) the anion concentration for dielectric BC increases monotonically from the boundary until the middle-point as shown in Fig.~\ref{fig:profiles-1V-1nm}(b) and (c). For metal BC, the stronger IC attraction leads to a peak of cation concentration close to the wall, as shown in Fig.~\ref{fig:profiles-1V-8nm}(b). Moreover, both cations and anions enrich near the surface for metal BC despite the positive applied potential.

Now, we turn to the case of a dielectric layer with dielectric constant $\epsilon_\text{L}$ on the metal surface that is lower than the dielectric constant of the solution. The dielectric layer generates a repulsive image charge that competes with the attractive image charge of the metal. This repulsive image charge is stronger with greater disparity between the layer and solution dielectric constant, and this effect is shown for a single planar boundary in Fig. \ref{fig:layer_singlewall}. We see that the ion density completely depletes from the surface, meaning the repulsive IC from the layer overpowers the metal IC at all distances. For a layer with a low dielectric constant $\epsilon_\text{L} = 2$, the repulsive IC is so strong that the ion profile nearly matches that of a dielectric wall with the same dielectric constant, shown by the dashed line in Fig. \ref{fig:layer_singlewall}. 

Next, we examine the double-layer structure between two parallel metal plates with a dielectric layer on each surface, shown in Fig. \ref{fig:layer_den}. For no applied potential and $\epsilon_\text{L} = 60$ in Fig. \ref{fig:layer_den}(a), the ions adsorb in the pore due to the metal's attractive IC. However, the ions deplete close to the wall due to the layer's repulsive IC, giving a peak in the ion density profile. The layer prevents ion saturation at the metal wall previously seen in Fig. \ref{fig:profiles-0V-1nm}, yielding density profiles that qualitatively resemble those of Ref.~\onlinecite{son2021image}. If the layer dielectric constant is low ($\epsilon_\text{L} = 2$), the ions completely deplete in the pore, similar to the profiles of the dielectric wall in Fig. \ref{fig:profiles-0V-1nm}, albeit with a higher maximum density due to the metal's attractive IC. For applied potential $V = 1$ in Fig. \ref{fig:layer_den}(b) and (c) at $\epsilon_\text{L} = 60$, the positive and negative ion density profiles are nonmonotonic for the same reasons as the zero-potential case.

\subsection{Differential capacitance}

We next examine the differential capacitance $C_d=d\sigma_s/dV$ for nanometer-sized slit pores. The two plates of the slit pore are  at the same surface potential $\psi(z=\pm L/2)= V$ to simulate a nanoscale pore inside a porous electrode that is maintained at a constant potential.

\begin{figure}[htbp!]
\centering
\includegraphics[width=0.8\columnwidth]{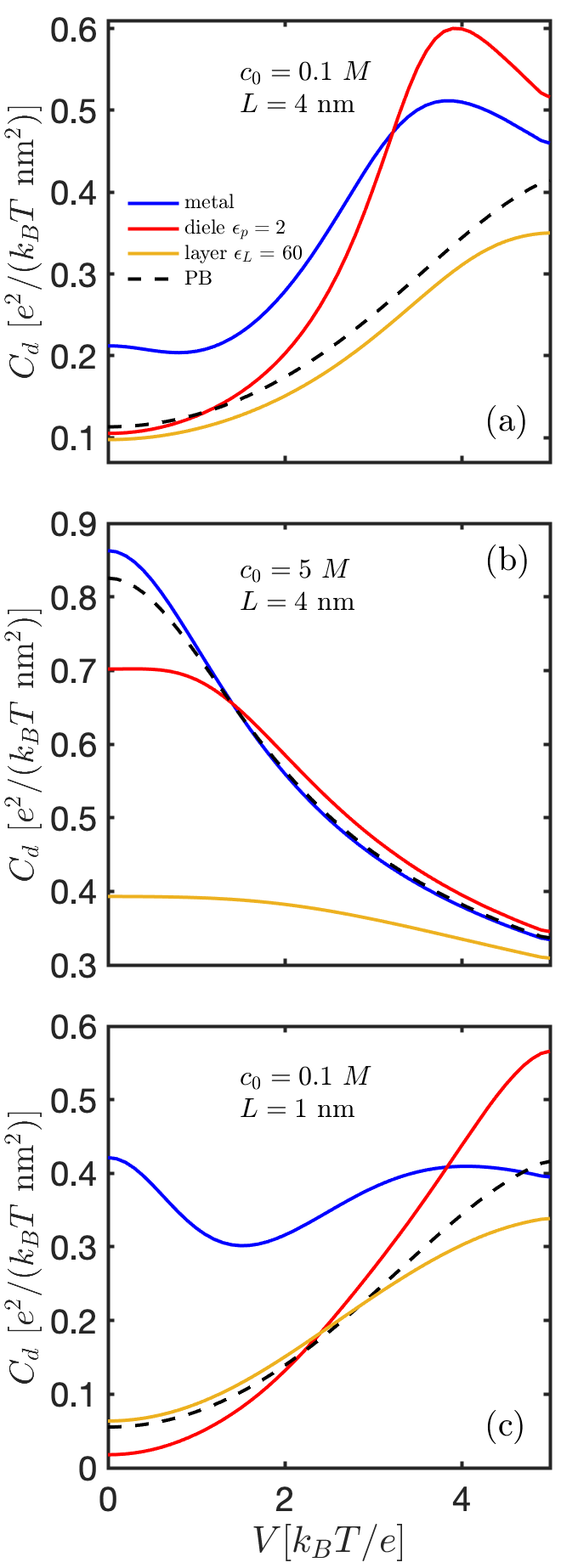}
\centering
\caption{
Differential capacitance curves for  four cases: metal (blue), dielectric (red), and layer BCs (yellow), as well as PB theory (black). The capacitance curves are shown under conditions:  
(a) $c_0=0.1$~M, $L=4$~nm.  
(b) $c_0=5$~M, $L=4$~nm.
(c) $c_0=0.1$~M, $L=1$~nm. 
\label{fig:capacitance}}
\end{figure}

\begin{figure}[htbp!]
\centering
\includegraphics[width=0.85\columnwidth]{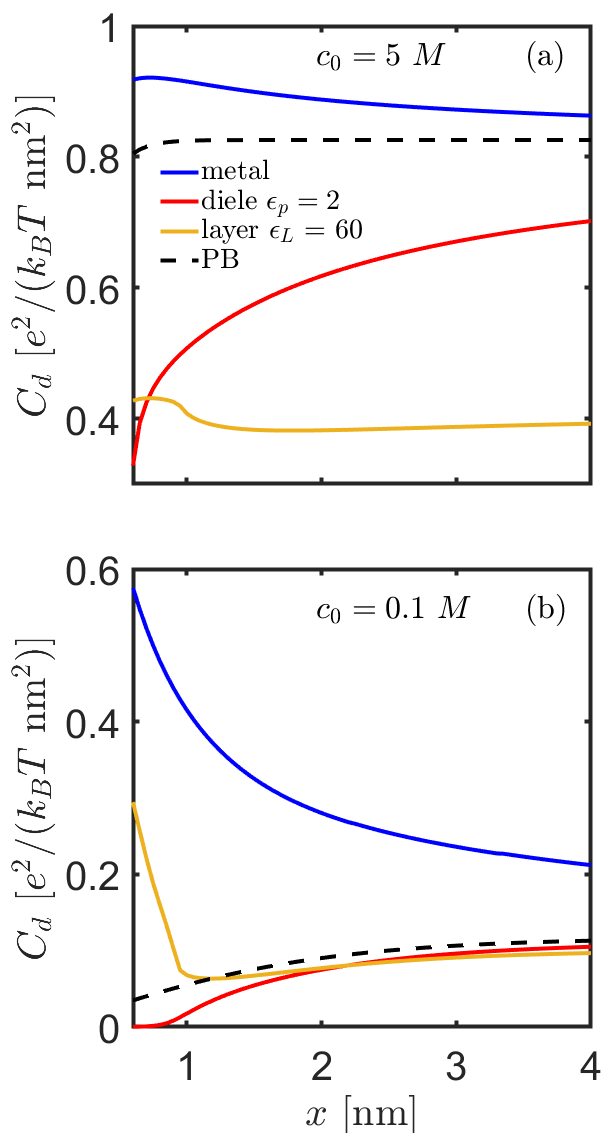}
\centering
\caption{
Differential capacitance curves for four cases: metal (blue), dielectric (red), and layer BCs (yellow), as well as PB theory (black) as a function of plate separation for $V = 0$. (a) Bulk ion concentration is $c_0 = 5$ M. (b) Bulk ion concentration is $c_0 = 0.1$ M.
\label{fig:cap_vs_L}}
\end{figure}

In all three system conditions explored in Fig. \ref{fig:capacitance}, the initial capacitance (at very low voltage) of the metal BC is higher than the PB theory, dielectric BC, or layer BC. This is consistent with the qualitative understanding that metal BC image charge attraction significantly enhances charge separation, where charge separation refers to the difference in anion and cation densities~\cite{son2021image}. On the other hand, at weak electric potentials, the image charge repulsion of dielectric and layer BCs strongly depletes salt ions and reduces charge separation. As a result, the initial $C_d$ is low. Moreover, since ions cannot closely approach the metal surface due to exclusion by the layer, the metal's surface charge is less sensitive to the applied potential. This effect, combined with the layer's repulsive image charge, gives a lower capacitance than the other three cases for nearly all conditions. The only exception is at low salt concentration, small separation, and low applied potential in Fig. \ref{fig:capacitance}(c), where the layer capacitance slightly exceeds dielectric BC and PB theory. The reason is that, for very small separations, the attractive IC from the metal dominates the layer's repulsive IC, causing greater accumulation of ions in the pore compared to dielectric BC and PB theory. 

At low bulk salt concentration $c_0=0.1$~M, and large separation between the plates ($L=4$~nm, compared to the double layer thickness $<1$~nm), the differential capacitance displays the ``camel-shape'' for dielectric and layer BCs, as well as PB theory~\cite{kornyshev2007double}, whereas that of the metal BC has a slight ``bird shape''~\cite{cruz2018bird}, as shown in Fig.~\ref{fig:capacitance}(a). For the PB theory, this is due to the nonlinear dependence of charge density on the electric potential. For dielectric BC, as surface potential increases, the attraction of the surface for counter-ions overcomes the image charge repulsion in the middle of the slit; therefore, the capacitance is enhanced. The same is true for the layer BC; however, since the metal's surface charge is less sensitive to the applied voltage for reasons mentioned earlier, the capacitance never exceeds the PB result except for small separations with low applied potential and low bulk concentration.

It is known that the overestimate of surface ion density in the Gouy--Chapman theory leads to a differential capacitance curve that is always ``camel-shaped'', with a minimum at 0 applied voltage. Kornyshev~\cite{kornyshev2007double} pointed out that the ``lattice saturation'' effect~\cite{eigen1954thermodynamics,kornyshev1981conductivity}, which accounts for finite ion sizes, can explain the observed deviations from the Gouy--Chapman theory, i.e. a ``bell'' shaped capacitance curve for room temperature ionic liquids~\cite{nanjundiah1997differential}.
Fig.~\ref{fig:capacitance}(b) shows this scenario with a high bulk salt concentration $c_0=5$~M, and large separation $L=4$~nm. The solvent concentration is only $c_s\approx3.33$~M in the bulk reservoir, lower than the ion concentrations, as in a `water-in-salt' electrolyte. At small voltage, initial charge separation is enhanced due to the large number of salt ions in the slit, giving a large $C_d$. Then, as the electric potential increases, the double layer saturates quickly, regardless of the BC. Thus, the overall $C_d$ displays a bell shape for all four cases. These results are consistent with previous studies~\cite{lockett2008differential,fedorov2008ionic,kornyshev2007double,qing2021surface}. Moreover, all curves converge at high applied potential due to double-layer saturation. We expect this convergence to occur for sufficiently high potential under the conditions for Fig. \ref{fig:capacitance}(a) and (c) as well. Similar to the modified PB theory, the ion-saturation effect plays a key role to render the bell shaped curves. 

The more interesting regime is at low bulk salt concentration $c_0=0.1$~M and very small separation $L=1.0$~nm. In this case, the boundary accumulation / depletion layers are very strong and dictate the charging behaviors at low surface potentials. Specifically, the IC attraction of metal BC strongly enhances the charge separation, resulting in a high initial capacitance. But, as the voltage increases, the saturated boundary layers screen out the influence of the surface potential in the middle of the slit, as seen in Fig.~\ref{fig:profiles-1V-1nm}. This leads to a small second peak of the blue solid line in Fig.~\ref{fig:capacitance}(c), giving a ``bird-shaped'' capacitance curve. On the other hand, for the dielectric and layer BCs, the IC repulsion at small distance is too strong at weak electric potential, so very few counter-ions migrate into the slit when the voltage increases initially. Hence, the low-voltage capacitance is quite low for the dielectric and layer BCs. Additionally, the linear response regime is much smaller for metal BC under low bulk salt and small separation compared to dielectric and layer BCs. The reason for this is that the charge fluctuation in the metal is much higher under these conditions, causing significant deviations from the linear response capacitance even at low applied potentials.

Lastly, we investigate the capacitance as a function of plate separation at $V = 0$ for two bulk ion concentrations $c_0 = 5$ and $0.1$ M, shown in Fig. \ref{fig:cap_vs_L} (a) and (b), respectively. At both concentrations, we see that the metal capacitance exceeds PB theory due to the attractive metal IC. Moreover, the metal capacitance increases as the pore size decreases, with an anomalously large increase for $c_0 = 0.1$ M compared to PB theory, similar to the results shown in Ref.~\cite{kondrat2011superionic}. For a dielectric boundary, the capacitance is lower than PB theory for both concentrations due to ion depletion in the pore caused by the dielectric wall's repulsive IC. For the layer BC in both cases, the capacitance is initially high due to the metal's attractive IC dominating at strong confinement. However, as the plates move apart, the capacitance decreases due to the layer's repulsive IC overtaking the attractive IC from the metal, causing ions to deplete in the pore. This effect creates a minimum in the capacitance. For large separations ($>1.5$) nm, the layer's repulsive IC has a weaker effect since it is a boundary layer effect, so the capacitance gradually increases.

\begin{figure}[htbp!]
\centering
\includegraphics[width=0.8\columnwidth]{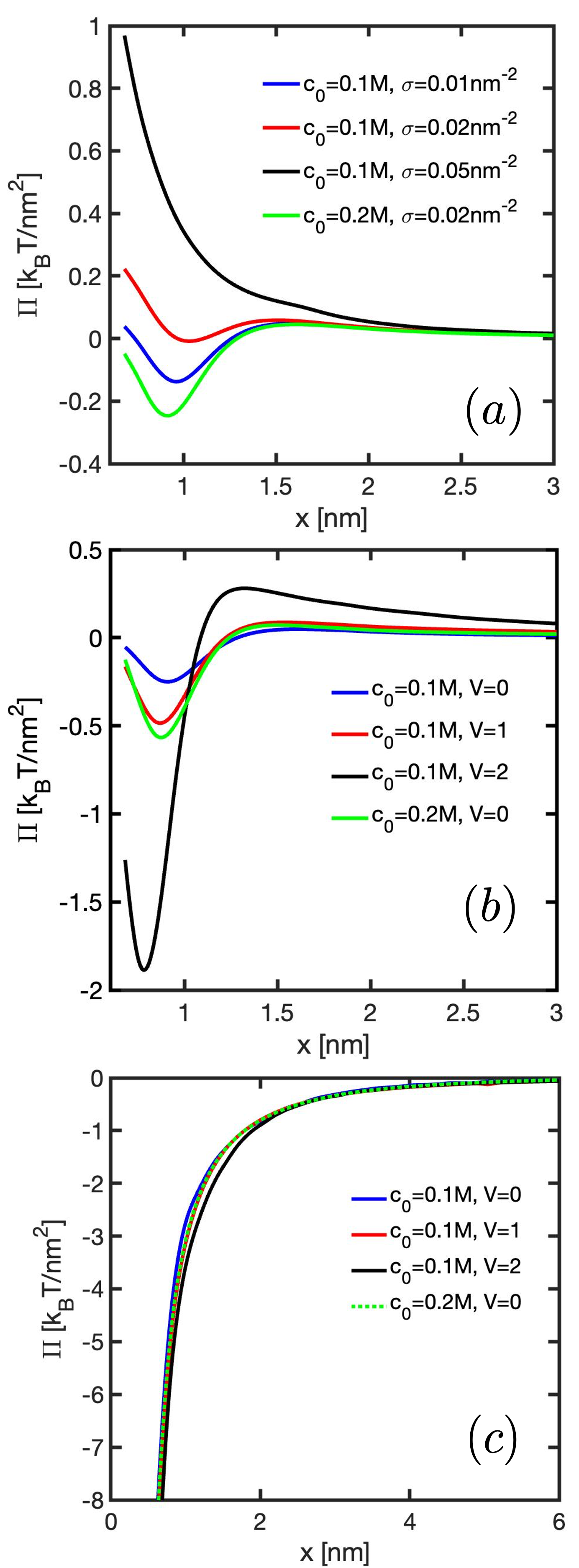}
\centering
\caption{
Force between neutral and like-charged plates with different material boundaries. In all plots the bulk concentrations are colored by: $c_0=0.1$~M (blue, red and black) and $c_0=0.2$~M (green).
(a) Two dielectric plates with same surface charge densities. Surface charge densities are $\sigma=0.01,0.02,0.05$~/nm$^2$ for the blue, red, and black line, respectively, and $\sigma=0.02$~/nm$^2$ for the green line.
(b) Two dielectric plates with the same fixed surface potentials. The surface potentials are $V=0,1,2$ for the blue, red, and black line, respectively, and $V=0$ for the green line.
(c) Two metal plates with the same fixed surface potentials. 
The surface potentials are $V=0,1,2$ for the blue, red, and black line, respectively, and $V=0$ for the green line.
\label{fig:force}}
\end{figure}

\begin{figure}[htbp!]
\centering
\includegraphics[width=0.8\columnwidth]{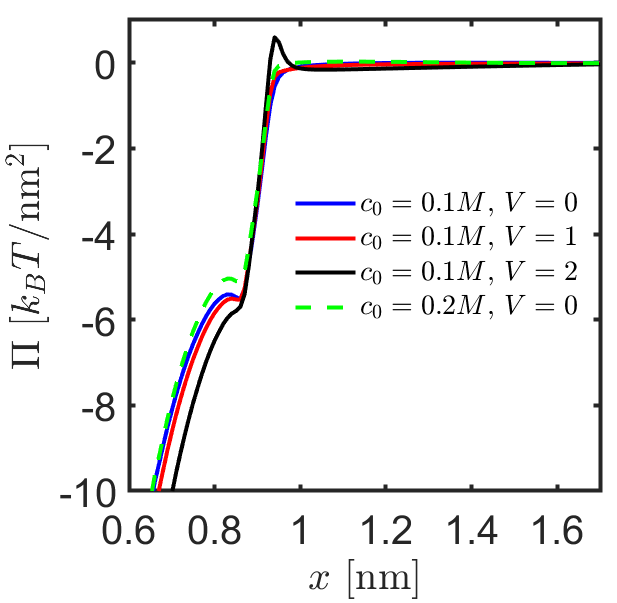}
\centering
\caption{
Force between neutral and like-charged metal plates with a $0.1$ nm thick dielectric layer with dielectric constant $\epsilon_\text{L} = 60$ on each plate. The metal plates have fixed surface potentials of either $V= 0,1,2$. The blue, red, and black curves have bulk ion concentration $c_0 = 0.1$ M, and the green curve has bulk ion concentration $c_0 = 0.2$ M.
\label{fig:force_layer}}
\end{figure}

\subsection{Force between like-charged plates}

Finally, we investigate the forces between two symmetrically charged plates. In reality, it is more natural to control surface charge density for  dielectric plates, or to maintain constant surface electric potential for metal plates, especially if the surface geometry is not simple. For ease of conceptual comparison, we also present results for dielectric BC with fixed surface potentials. In Fig.~\ref{fig:force} (a), we investigate three surface charges $\sigma=0.01, 0.02, 0.05$~e/nm$^2$ at bulk concentration $c_0 = 0.1$ M (blue, black, and red curves, respectively). For lower surface charges $\sigma=0.01, 0.02$~e/nm$^2$, the ion depletion due to IC of dielectric BC results in an attractive potential well. There is also a small but noticeable repulsive regime  for plate separations between 1$\sim$2~nm, which is consistent with the findings of Ref.~\cite{wang2013effects}, although there it was computed using a WKB approximation. At higher surface charge $\sigma=0.05$~e/nm$^2$, the same-charge repulsion overcomes the IC effect. Moreover, for $c_0=0.1$~M, the attractive well gradually disappears and completely vanishes when increasing the surface charge from $0.01$ to $0.05$~e/nm$^2$ in Fig.~\ref{fig:force} (a). Moving to the effect of bulk concentration, the green curve is for a fixed surface charge $\sigma=0.02$~e/nm$^2$ with bulk concentration $c_0 = 0.2$ M. Comparing the green curve to the red curve, it is obvious that the higher bulk concentration leads to a stronger depletion force at the same surface charge density. This is also qualitatively consistent with Ref.~\cite{wang2013effects}. However, when the boundary condition is changed to fixed surface potentials, the purely repulsive regime is no longer observed for reasonably weak applied potentials, as shown in Fig.~\ref{fig:force}(b). The repulsion at short distance in Fig. \ref{fig:force}(a) comes from the strong repulsion between the fixed surface charges, while in Fig. \ref{fig:force}(b) the surface charges  adjust to avoid strong repulsion and to lower the total free energy. At fixed surface potential, both higher surface potential (black line) and higher bulk salt concentration (green line) amplify the  variation in the force, as the IC-induced depletion effect is enhanced.

In the case of metal BC, the IC attraction is very strong and dominates the interplate force. This is shown by Fig.~\ref{fig:force}(c), where the magnitude of the attraction is much larger than in Fig. \ref{fig:force}(b). In addition, no repulsive regime is observed and the  attractive force decays over a longer lengthscale.  Recall that we use point-charge model throughout the calculations here for consistency and fair comparisons. This means that ions can approach the boundaries infinitely close. For dielectric BC, the IC repulsion self-regulates this effect by suppressing the ion distributions at the surface. However, for metal BC, the IC attraction results in high concentration of counter-ions at the surface close to the saturation concentration, with a very large self-energy. As a result the attraction force computed here at short separations between the two electrodes may be overestimated compared with experimental or simulation results with finite ion sizes. For this reason, we introduce the thin dielectric layer to the metal surfaces to account for the distance of closest approach for ions, which we discuss next.

To finish, we investigate the effect of a $0.1$ nm thick dielectric layer (dielectric constant $\epsilon_\text{L} = 60$) on metal plates, where the metal plates are at the same fixed potential. In Fig. \ref{fig:force_layer}, we see that the layer's repulsive IC cancels the metal's attractive IC for separations $>1$ nm, preventing the plates from interacting until close separation ($<1$) nm for all conditions considered. Consequently, the lengthscale of interaction between plates with layer BC is much smaller than that of metal and dielectric BCs. Moreover, at separations of $1$ nm and above, the ions are depleted from the walls. As the plates move from 1 nm to 0.9 nm spacing, the metal's attractive IC overcomes the repulsive IC from the layer, causing ions to flood into the pore that generates an attractive force between the plates. Furthermore, the peak in the $V = 2$ curve stems from a stronger flooding of ions into the pore compared to the lower potentials. The slight bump in the force curves near 0.85 nm separation is due to the ions saturating the pore walls.

\section{Conclusions}

In this work, we discuss the image charge effect
and its consequences in nano-capacitors and forces between like-charged plates. We treat the IC effects non-perturbatively for a point-charge model, based on a renormalized Gaussian fluctuation theory. We apply the incompressibility condition to account for the excluded volume interactions between the ions and solvent and to prevent the ion density from diverging at the interfaces. We obtain full numerical solutions to the renormalized field theory subject to different boundary conditions. We focus on a comparison of these calculations under the (1) perfectly conducting metal plates, (2) dielectric plates, (3) metal plates with a thin dielectric layer, and (4) the PB theory. We demonstrate these comparisons using a simple 1:1 electrolyte.

We first show that the double-layer structures are significantly affected by the IC interaction. For a single surface without applied voltage or externally-imposed surface charge, IC leads to saturation of ions at the surface for the metal BC, and strong ion depletion for the dielectric and layer BCs. For the layer BC, the depletion strengthens with lower dielectric constant relative to the solvent. For a narrow slit pore, the middle-point concentrations significantly deviate from the bulk values due to IC. When a single surface is fixed at a weak applied potential, the electrostatic potential profiles are not sensitive to the BCs. However, ion densities near the surface depends strongly on the type of BC applied. For a narrow slit pore with both surfaces at a weak fixed potential, both the ion density and the electrostatic potential profiles are strongly affected by the BCs. Moreover, the layer BC gives nonmonotonic ion density profiles that are qualitatively similar to those observed in MD simulations~\cite{son2021image}.

Based on the understanding about the double-layer structures, our calculations further show that the boundary layers due to IC interaction lead to significant differences in the differential capacitance of a nano-pore under different boundary conditions. The dielectric and layer BCs show a qualitatively similar trend with the PB theory, where the capacitance curve changes from a camel shape to a bell shape as bulk salt concentration increases.  Moreover, the metal BC gives a ``bird-shaped'' capacitance at low bulk ion concentrations and small pore width due to IC-induced boundary layer accumulation and ion saturation. The layer BC shows that the ion exclusion near the metal surface combined with the repulsive IC yields a lower capacitance than PB, metal, or dielectric BCs at nearly all conditions.
Notice that modified PB treatment~\cite{kondrat2011superionic} only accounts for the IC from the mean-charge density $\rho=c_+-c_-$ in the Poisson equation, while our treatment captures the IC effect even at $\rho=0$, and the ion density profiles $c_+$ and $c_-$ are non-uniform. As a result, our predictions of the capacitance are different from the modified PB treatment, especially at weak applied potential.

The differences between the BCs are also manifested in the forces between like-charged planar walls. For dielectric BC, at fixed surface charge density, the attraction well gradually diminishes as surface charge increases; at fixed surface potential, the attraction well is stronger at higher potential values. For metal BC with fixed surface potentials on both plates, which is more directly relevant to nano-pores inside porous electrodes, the attraction is much stronger than the dielectric BC and decays slower; no repulsion is observed in this ensemble in the weak-coupling regime. Increasing either surface potential or bulk concentration slightly enhances the attraction force. The prediction of pure attraction for planar metal plates is different from the nonlinear PB theory, which only gives repulsion between like-charged metal plates---Ref.~\onlinecite{dos2019like} showed that in the PB theory, one has to invoke curvature effects to induce like-charge attraction with metal BC. For layer BC at fixed surface potentials, the interaction lengthscale is much smaller than that of metal or dielectric BCs due to the layer's repulsive IC canceling the metal's attractive IC at separations $>1$ nm. For separations $<1$ nm, the plates are strongly attractive, similar to the metal BC, due to ions flooding the pore at close separation when the metal IC overcomes the layer's repulsive IC.

Our work presents a fundamental perspective on the IC interaction, highlighting its non-negligible influence on double layer structures, and consequently, nanoscale capacitors and forces between like-charged plates. These effects are often omitted or treated in a mean-field fashion. However, we show that even at zero mean-charge density, where mean-field predictions will neglect IC, it still has a strong effect. Our results provide a starting point to include IC in various realistic applications.

\begin{acknowledgments}
T.Z. acknowledges the support of the Cecil and Sally Drinkward Postdoc Fellowship. He also thanks Z. Peng for helpful discussions on numerical methods. Z.-G.W. acknowledges financial support from the Hong Kong Quantum AI Lab Ltd.

\end{acknowledgments}

\section*{Data Availability Statement}
All data that support the findings of this study are available within the article.

\appendix

\section{Brief summary of the key equations}

Most of the derivations were already presented in Ref.~\onlinecite{wang2010fluctuation} and earlier references cited there. Here we sketch the outline of the derivations, including the incompressibility constraint, for the reader's convenience.
We start with the Hamiltonian in SI unit
\begin{equation}
H = \int d\br d\br' \frac{e^2}{2} \rho(\br) G_0(\br,\br') \rho(\br')
\end{equation}
where the Coulomb operator is defined by 
\begin{equation}
- \nabla\cdot \varepsilon \nabla' G_0(\br,\br') = \delta(\br-\br')
\end{equation}
The total charge density can be decomposed into external charge, cations and anions
\begin{equation}
    \rho(\br) = \rho_{ex}(\br) + z_+ \sum_{i+} h_+(\br-\br_+^i) - z_- \sum_{j-} h_-(\br-\br_-^j)
\end{equation}
where $h_{\pm}$ represents the spread shape of ion charge.
The incompressibility of the liquid requires 
\begin{equation}
\hat{c}_{tot} = \hat{c}_+(\br) + \hat{c}_-(\br) + \hat{c}_s(\br) = 1/v_0
\end{equation}
where the density of species
\begin{equation}
\begin{split}
\hat{c}_+(\br) = & \sum_{i+} h_+(\br-\br_+^i) \\
\hat{c}_-(\br) = & \sum_{i-} h_-(\br-\br_-^i) \\
\hat{c}_s(\br) = & \sum_{s-} h_s(\br-\br_s^i)
\end{split}
\end{equation}
subscript $s$ stands for solvent.

The canonical partition function is
\begin{equation}
\begin{split}
Q = & \frac{1}{n_+!n_-!n_s! v_+^{n_+}v_-^{n_-}v_s^{n_s}}\int d\br_{i+} d\br_{j-} d\br_{s} e^{-\beta H} \prod_{r} \delta\left(\hat{c}_{tot} - 1/v_0\right) \\
= & \frac{1}{n_+!n_-!n_s! v_+^{n_+}v_-^{n_-}v_s^{n_s}}\int d\br_{i+} d\br_{j-} d\br_{s}\\ 
& \times e^{-\beta \frac{e^2}{2} 
\int d\br d\br' (\rho_{ex} + \rho_{ion})(\br) G_0(\br,\br') (\rho_{ex} + \rho_{ion})(\br')
}\\
& \times\int dc_{\pm,s}\int D[\eta] e^{ i \int d\br \eta(\br) ({c}_+ + {c}_- + {c}_s - 1/v_0)(\br)}\\
& \times \int dw_{\pm,s} e^{
i \int d\br \left[
w_+ (\hat{c}_{+} - c_{+}) +w_- (\hat{c}_{-} - c_{-}) + w_s (\hat{c}_{s} - c_{s})
\right]}
\end{split}
\end{equation}
where auxiliary fields $w_{\pm,s}$ are also introduced. In general, if there are other quadratic or higher-order interaction terms involving the concentrations, it will be useful to deal with the $w$ fields instead of directly manipulating the concentration operators.
Using the Hubbard-Stratonovich transform
\begin{equation}
    \int dx \exp\left( -\frac{1}{2} x^T A^{-1} x \pm ixy \right) = \sqrt{\det(A)} \exp \left( -\frac{1}{2} y^T A y \right)
\end{equation}
the exponential factor is 
\begin{equation}
\begin{split}
& e^{-\beta \frac{e^2}{2} 
\int d\br d\br' \rho(\br) G_0(\br,\br') \rho(\br')
} \\
& = \frac{1}{\sqrt{\det(G_0)}} \int D[\phi] e^{
-\int d\br d\br' \left( \frac{1}{8 \pi \ell_B} \nabla \phi(\br)\cdot \nabla' \phi(\br') + i \rho(\br) \phi(\br')
\right)\delta(\br-\br')} \\
\end{split}
\end{equation}
here $\ell_B=e^2/(4\pi\epsilon k_B T)$ is the Bjerrum length and $\phi$ is the dimensionless potential scaled by $\frac{k_B T}{e}$. 

Now the grand partition function is
\begin{equation}
\begin{split}
\Xi = & \sum_{n_{+}}\sum_{n_-}\sum_{n_s} \frac{Q e^{n_+\mu_+} e^{n_-\mu_-} e^{n_s\mu_s}}{\sqrt{\det(G_0)}} \int D[\phi]D[\eta]dw_{\pm,s}dc_{\pm,s} e^{-\mathcal{L}}
\end{split}
\end{equation}
where the action is
\begin{equation}
\begin{split}
\mathcal{L} = &
\int d\br d\br' \left[ \frac{1}{8 \pi \ell_B} \nabla \phi(\br)\cdot \nabla' \phi(\br') + i \rho_{ex}(\br) \phi\br')
\right]\delta(\br-\br')  \\
& - \int d\br \left( \lambda_+ e^{- i\left(h_+*[z_+\phi+iw_+]\right)(\br)} + \lambda_- e^{ i \left(h_-*[z_-\phi-iw_-]\right)(\br)}\right) \\
& + \int d\br \left( w_+ c_+ + w_- c_- + w_s c_s  - \eta (c_+ + c_- + c_s - 1/v_0) \right)\\
&- \int d\br \lambda_s e^{ h_s*w_s}
\end{split}
\end{equation}
$*$ stands for the convolution operator
\begin{equation}
(h*\phi)(\br) = \int d\br' h(\br'-\br)\phi(\br')
\end{equation}
and $\lambda_{\pm} = \frac{e^{\mu_{\pm}}}{v_{\pm}}$ is the fugacity.

The variational approach decomposes the field $\phi=-i\psi+\chi$, and the action becomes
\begin{equation}
\begin{split}
\mathcal{L}[\phi]
& = \int d\br \left\{ \frac{1}{8 \pi \ell_B} \left(\nabla \left(-i\psi(\br)+\chi(\br)\right)\right)^2  + \rho_{ex}(\br) (-i\psi(\br)+\chi(\br))\right\}
\\
& - \int d\br \lambda_{+} e^{-z_{+}(h_{+}*\psi)(\br)+(h_{+}*w_+)(\br) -i z_{+}(h_{+}*\chi)(\br)} \\
& - \int d\br \lambda_{-} e^{z_{-}(h_{-}*\psi)(\br)+(h_-*w_-)(\br) + i z_{-}(h_{-}*\chi)(\br)}\\
& + \int d\br \left( w_+ c_+ + w_- c_- + w_s c_s  - \eta (c_+ + c_- + c_s - 1/v_0) \right)\\
&- \int d\br \lambda_s e^{ h_s*w_s}
\end{split}
\end{equation}
The average number (density) of ions is
\begin{equation}
\begin{split}
\left< n_{\pm} \right> =  \frac{\partial \ln\Xi}{\partial \mu_{\pm}} = & \int d\br \lambda_{\pm} \left< e^{\mp z_{\pm}(h_{\pm}*(\psi+i\chi))(\br)+(h_{\pm}*w_{\pm})(\br)} \right> \\
c_{\pm}(\br) = & \lambda_{\pm} \left< e^{\mp z_{\pm}(h_{\pm}*(\psi+i\chi))(\br)+(h_{\pm}*w_{\pm})(\br)} \right>
\end{split}
\end{equation}

For the variatioal renormalization, a Gaussian reference action is chosen
\begin{equation}
\mathcal{L}_{ref} = \int d\br d\br' \frac{1}{2} \chi(\br) G^{-1}(\br,\br') \chi(\br') 
\end{equation}
and the extremized free energy is
\begin{equation}
\begin{split}
W = & W_{ref} + \left< \mathcal{L}[\phi] - \mathcal{L}_{ref}[\phi] \right>_{ref} \\
= & -\frac{1}{2} \ln\left( \frac{\det(G)}{\det(G_0)}\right) + \left< L[\phi] \right>_{ref} - \frac{1}{2}
\end{split}
\end{equation}
To evaluate $\left< \mathcal{L}[\phi] \right>_{ref}$, we note that
\begin{equation}
\begin{split}
& \left< \int d\br d\br' \frac{1}{4\pi \ell_B(\textbf{r})}\delta(\br-\br') \nabla\chi(\br) \cdot \nabla' \chi(\br') \right>_{ref} \\
= &  \int d\br d\br' \left\{ \left<\chi(\br) \chi(\br')\right>_{ref} \nabla\left[\frac{1}{4\pi \ell_B(\textbf{r})}\right] \cdot \nabla'\delta(\br-\br') \right\}  \\
= & \int d\br d\br' \left\{
G(\br,\br') \nabla\left[\frac{1}{4\pi \ell_B(\textbf{r})}\right] \cdot \nabla'\delta(\br-\br') \right\}  \\
\end{split}
\end{equation}
and
\begin{equation}
\begin{split}
\left< e^{\mp i z_{\pm}
(h_{\pm}*\chi)(\br)} \right>_{ref}
= &
e^{
-\frac{1}{2} z_{\pm}^2 \int d\by d\by' h_{\pm} (\by-\br) G(\by,\by') h_{\pm}(\by'-\br)
}
\end{split}
\end{equation}
Now we can simplify the averaged action as
\begin{multline}
\left<\mathcal{L}[\psi]\right>_{ref} = \int dr\left\{ -\frac{1}{8 \pi \ell_B}(\nabla\psi(\br))^2 + \rho_{ex}(\br)\psi(\br) - c_{\pm}(\br) \right\} \\
 + \frac{1}{2} \int dr dr' G(\br,\br') \nabla \cdot \left[\frac{1}{4\pi \ell_B(\textbf{r})}\right]\nabla'\delta(\br-\br') \\
 + \int d\br \left( w_+ c_+ + w_- c_- + w_s c_s  - \eta (c_+ + c_- + c_s - 1/v_0) \right)\\
- \int d\br \lambda_s e^{ h_s*w_s}
\end{multline}

The Euler-Lagrange equations in the main text for this field theory then result from variational derivatives of $W$ with respect to $\psi$, $G$, $w_{\pm,s}$, $c_{\pm,s}$ and $\eta$.

\section{Image charge of a single point charge near layered interfaces}

\subsection{Metal-dielectric interface}
Suppose for half space $z<0$, it's a metal plate. For half space $z>a$, it's a dielectric medium with relative permitivity $\epsilon_2$. In between $0<z<a$ it's another dielectric with $\epsilon_1$. A point charge $q$ is placed at $(0,0,a+b)$. We now solve for the potential inside the two dielectric media $\phi_1$ and $\phi_2+\phi_3$.
Due to cylindrical symmetry, the Poisson equation can be written as
\begin{equation}
\begin{split}
&\left(
\partial_z^2 + \frac{1}{r}\partial_r r \partial_r + \frac{1}{r^2}\partial_{\theta}^2
\right)\phi_i = 0 \qquad i=1,2\\
& \phi_3 = \frac{eq}{4\pi\epsilon_0\epsilon_2\sqrt{r^2+[z-(a+b)]^2}}
\end{split}
\end{equation}
Nondimensionalize by the Bjerrum length in the liquid
\begin{equation}
\ell_B = \frac{e^2}{4\pi \epsilon_0 \epsilon_2 k_B T}
\end{equation}
\begin{equation}
\psi = \frac{e\phi}{k_B T} = \frac{q \ell_B}{\sqrt{r^2+[z-(a+b)]^2}}
\end{equation}
For water $\epsilon_2=80$, $\ell_B\sim0.7$~nm.

On the metal-dielectric interface $z=a$, the boundary conditions are
\begin{equation}
\label{eqn:BC-a}
\begin{split}
\epsilon_1 \partial_z \phi_1(r,\theta, a) & = \epsilon_2 \partial_z \left(\phi_2(r,\theta, a)+\phi_3(r,\theta, a)\right)\\
\phi_1(r,\theta, a) & = \phi_2(r,\theta, a)+\phi_3(r,\theta, a)\\
\end{split}
\end{equation}
the second condition implies already the continuity of the tangential $E$ field
\begin{equation}
\partial_r \phi_1(r,\theta, a) = \partial_r \left( \phi_2(r,\theta, a)+\phi_3(r,\theta, a)\right)
\end{equation}
At $z=0$,
\begin{equation}
\phi_1(r,\theta,0)=0
\label{eqn:BC-0}
\end{equation}
At $z\rightarrow\infty$,
\begin{equation}
\phi_2(r,\theta,\infty)=0
\label{eqn:BC-inf}
\end{equation}
The general solution satifying BCs Eqn.~\ref{eqn:BC-0} and Eqn.~\ref{eqn:BC-inf}
\begin{equation}
\begin{split}
\phi_1 & = \int_0^{\infty} d\lambda \sinh(\lambda z) J_0(\lambda r) f_1(\lambda)\\
\phi_2 & = \int_0^{\infty} d\lambda
e^{-\lambda z} J_0(\lambda r) f_2(\lambda)
\end{split}
\end{equation}
Then one use the fact 
\begin{equation}
\frac{1}{\sqrt{r^2+(z-a-b)^2}} = \int_0^{\infty} d\lambda J_0(\lambda r) e^{-\lambda\vert z-a-b\vert}
\end{equation}
To match the BC at $z=a$
\begin{equation}
\begin{split}
q\epsilon_2 e^{-\lambda b} & = f_1(\lambda) \epsilon_1 \cosh(a \lambda) + f_2(\lambda) \epsilon_2 e^{-\lambda a} \\
q e^{-\lambda b} & = f_1(\lambda)  \sinh(a \lambda) - f_2(\lambda) e^{-\lambda a}
\end{split}
\end{equation}
As a result,
\begin{equation}
\begin{split}
f_1(\lambda) & = \frac{2\epsilon_2 q e^{-\lambda b}}{\epsilon_1\cosh(\lambda a)+ \epsilon_2\sinh(\lambda a)}\\
f_2(\lambda) & = q\frac{ \epsilon_2\sinh(\lambda a)-\epsilon_1\cosh(\lambda a)}{\epsilon_1\cosh(\lambda a)+ \epsilon_2\sinh(\lambda a)}e^{\lambda (a-b)}\\
\end{split}
\end{equation}

For the IC force, at $z=a_{+0}$, the excess E field is
\begin{equation}
\begin{split}
E_r &= -\int_0^{\infty} d\lambda
e^{-\lambda a} \lambda J_0'(\lambda r) f_2(\lambda)\\
& = \int_0^{\infty} d\lambda
e^{-\lambda a} \lambda J_1(\lambda r) f_2(\lambda)\\
E_z &= \int_0^{\infty} d\lambda
e^{-\lambda a} \lambda J_0(\lambda r) f_2(\lambda)
\end{split}
\end{equation}
in addition to the E field generated by the point charge potential $\phi_3$.

The surface charge density $\sigma(z=a)$ on the interface at $z=a$ can be extracted by the difference of the E field
\begin{equation}
\begin{split}
E_z(z=a+0)-E_z(z=a-0) = \sigma(z=a)\left(\frac{1}{\epsilon_1} + \frac{1}{\epsilon_2}
\right)
\end{split}
\end{equation}
Similarly, for the surface charge density $\sigma(z=0)$, we have
\begin{equation}
\sigma(z=0) = \epsilon_1 E_z(z=0+0)
\end{equation}

\section{ IC with a dielectric surface layer} \label{appendix:layer}
Suppose $z>0$ half space is filled by the electrolyte solution with dielectric constant $\epsilon_r$, while $-a<z<0$ is a dielectric layer with $\epsilon_\text{L}$. 

The BC for Poisson equation with fixed surface potentials is a Robin type
\begin{equation}
\epsilon_\text{L} \phi(0) - \epsilon_r a \phi'(0) = \epsilon_\text{L} V_1
\end{equation}
For another layer at $L<z<L+a$, we have similarly
\begin{equation}
\epsilon_\text{L} \phi(L) + \epsilon_r a \phi'(L)  = \epsilon_\text{L} V_2
\end{equation}

For fixed surface charge $e\sigma_S$ at $z=-a$, the BC at $z=0$ is still a Neumann BC
\begin{equation}
\begin{split}
e\sigma_S = -\epsilon_\text{L} \phi_\text{L}'(0) = -\epsilon_r \phi_r'(0) \end{split}
\end{equation}

For the self-energy Green's function, a Robin type BC now applies to both metal and dielectric plates at $z<-a$. For a dielectric plate with $\epsilon_p$
\begin{equation}
G'(0) - k \frac{\epsilon_\text{L}}{\epsilon_r}\frac{\epsilon_\text{L} \sinh(ka) + \epsilon_p \cosh(ka)}{\epsilon_p \sinh(ka) + \epsilon_\text{L} \cosh(ka)} G(0) = 0
\end{equation}
Taking the limit of $\epsilon_p\rightarrow\infty$, we have the BC for the metal plate
\begin{equation}
G'(0) - k\frac{\epsilon_\text{L}}{\epsilon_r}\coth(ka) G(0) = 0
\end{equation}



\nocite{*}
\bibliography{imagecharge}

\end{document}